\documentclass[twocolumn,showpacs,amsmath,amssymb,bibnotes]{revtex4}
\usepackage{bm}
\usepackage{graphicx,psfig,epsfig,subfigure,afterpage}
\usepackage{amsfonts}

\newcommand{\vect}[1]{\underline{#1}}
\newcommand{\tens}[1]{\underline{\underline{#1}}}

\newcommand{\vm}{\vect{v}_{\,\rm m}}
\newcommand{\vs}{\vect{v}_{\,\rm s}}

\newcommand{\vrel}{\vect{v}_{\, \rm rel}}
\newcommand{\vv}{\vect{v}}

\newcommand{\etam}{\eta_{\,\rm m}}
\newcommand{\etas}{\eta_{\,\rm s}}

\newcommand{\Dm}{\tens{D}_{\,\rm m}}
\newcommand{\Ds}{\tens{D}_{\,\rm s}}

\newcommand{\Omm}{\tens{\Omega}_{\,\rm m}}

\newcommand{\nablu}{\vect{\nabla}}

\newcommand{\be}{\begin{equation}}
\newcommand{\ee}{\end{equation}}
\newcommand{\bea}{\begin{eqnarray}}
\newcommand{\eea}{\end{eqnarray}}

\newcommand{\gdot}{\dot{\gamma}}

\newcommand{\gae}{\stackrel{>}{\scriptstyle\sim}}
\newcommand{\lae}{\stackrel{<}{\scriptstyle\sim}}
\newcommand{\ie}{{\it i.e.\/}}
\newcommand{\eg}{{\it e.g.\/}}

\newcommand{\bw}{\begin{widetext}}
\newcommand{\ew}{\end{widetext}}

\newcommand{\bmini}{\begin{minipage}}
\newcommand{\emini}{\end{minipage}}

\newcommand{\eigenvec}{\vect{{\tt{v}}}_{\vect{k},\alpha}}

\begin{document}

\title{Early stages of the shear banding instability in wormlike micelles}
\author{S. M. Fielding}
\email{physf@irc.leeds.ac.uk} 
\author{P. D. Olmsted}
\email{p.d.olmsted@leeds.ac.uk}
\affiliation{Polymer IRC and
  Department of Physics \& Astronomy, University of Leeds, Leeds LS2
  9JT, United Kingdom} 
\date{\today} 
\begin{abstract}
  We study the early stages of the shear banding instability in
  semidilute wormlike micelles using the non-local Johnson-Segalman
  model with a two-fluid coupling of the concentration $\phi$ to the
  shear rate $\gdot$ and micellar strain $\tens{W}$.  We calculate the
  ``spinodal'' limit of stability for sweeps along the homogeneous
  intrinsic flow curve.  For startup ``quenches'' into the unstable
  region, the instability in general occurs before the homogeneous
  startup flow can attain the intrinsic flow curve. We predict the
  selected time and length scales at which inhomogeneity first
  emerges.  In the ``infinite drag'' limit, fluctuations in the
  mechanical variables $\gdot$ and $\tens{W}$ are independent of those
  in $\phi$, and are unstable when the slope of the intrinsic flow
  curve is negative; but no length scale is selected.  For finite
  drag, the mechanical instability is enhanced by coupling to $\phi$
  and a length scale is selected, in qualitative agreement with recent
  experiments.  For systems far from an underlying zero-shear demixing
  instability this enhancement is slight, while close to demixing the
  instability sets in at low shear rates and is essentially demixing
  triggered by flow.
\end{abstract}
\pacs{{47.50.+d}{ Non-Newtonian fluid flows}--
     {47.20.-k}{ Hydrodynamic stability}--
     {36.20.-r}{ Macromolecules and polymer molecules}
}
\maketitle
\begin{figure}[h]
\begin{center}
\includegraphics[scale=0.25]{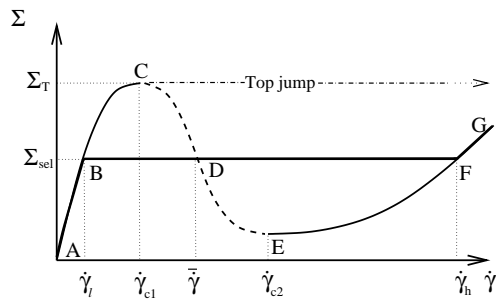}
\vspace{-0.4cm} 
\caption{Schematic flow curve.
\label{fig:schem} } 
\vspace{-0.3cm} 
\end{center}
\end{figure}

Control of morphology and stability is vital for processing many
complex fluids, \eg\ polymeric, liquid crystalline, and surfactant
fluids. While much is understood close (or relaxing) to equilibrium,
strongly driven systems suffer in comparison. For many complex fluids,
the intrinsic constitutive curve of shear stress $\Sigma$ versus shear
rate $\gdot$ is non-monotonic.  For semi-dilute wormlike micelles,
theory predicts the form ACEG of
Fig.~\ref{fig:schem}~\cite{cates90,SpenCate94}.  In the regime of
decreasing stress, $\gdot_{\rm c1}<\gdot<\gdot_{\rm c2}$, homogeneous
flow is unstable~\cite{Yerushalmi70,SCM93} and the system splits into
bands of different shear rates $\gdot=\gdot_{\ell},\gdot_{\rm h}$
~\cite{rehage,SANS,FB,berret94b,Call+96},
with a steady state flow curve ABFG.  Constitutive models augmented with
interfacial gradient terms have captured this
behaviour~\cite{olmsted99a,Dhon99}.
In this paper we study the linear stability of initially
homogeneous states.
We also predict the length and time scales at which inhomogeneity
emerges after a startup ``quench'', \ie\ imposing a shear rate in the
unstable regime. Our analysis is the counterpart, for this {\em
  driven} phase transition, of the Cahn-Hilliard (CH) calculation for
demixing in {\em undriven} systems.

Shear startup 
experiments~\cite{berret94b,Berr97,DecLerBer01} reveal (i) a
metastable regime $\gdot_\ell<\gdot\lae\gdot_{\rm c1}$ of slow
approach to the banded state, and (ii) an unstable regime, onset at
$\gdot\gae \gdot_{\rm c1}$, where the stress can massively overshoot
$\Sigma_{\rm sel}$ before subsiding rapidly to it. In
Ref.~\cite{DecLerBer01}, the overshoot coincided with concentration
fluctuations that emerged perpendicular to the shear compression axis
at a selected length scale $O(1\mu \rm{m})$. The authors attributed
this to the Helfand-Fredrickson (HF)~\cite{HelfFred89} coupling of
flow to concentration. This features in two component systems with
well separated relaxation times (\eg\ polymer and
solvent)~\cite{brochdgen77,HelfFred89,DoiOnuk92,milner93,WPD91,ClarMcle98}.
The slow component (polymer) tends to migrate to regions of high
stress. If the plateau modulus increases with polymer concentration,
positive feedback enhances concentration fluctuations and can shift
the spinodal of any nearby CH instability\cite{ClarMcle98}.

Here we model the initial stage of the unstable kinetics using the
non-local Johnson-Segalman (d-JS) model~\cite{johnson77} within a
2-fluid framework~\cite{brochdgen77,milner93} that incorporates
concentration dynamics.  The 2-fluid model considers separate force
balance equations for the micelles (velocity $\vm$, volume fraction
$\phi$) and solvent (velocity $\vs$).  These are added to give overall
force balance for the average velocity $\vv=\phi\vm+(1-\phi)\vs$
(Eq.~\ref{eqn:navier}) and subtracted for the relative velocity
$\vrel=\vm-\vs$, which in turn specifies the concentration
fluctuations (Eq.~\ref{eqn:relative}): \vspace{-0.6cm}
\begin{widetext}
\vspace{-0.4cm}
  \begin{gather}
\rho \left(\partial_t+\vect{v}.\nablu\right)\vv\equiv D_t\vect{v}
=\nablu.G(\phi)\,\tens{W}-\phi\,\nablu\frac{\delta
  F(\phi)}{\delta\phi}+2\,\nablu.\,\phi\, \etam\,
\Dm^{0}+2\nablu.\,(1-\phi)\, \etas\, \Ds^{0} -\nablu p,     
\label{eqn:navier}\\
D_t\phi=-\nablu\cdot\phi(1-\phi)
\vrel=-\nablu\cdot\frac{\phi^2(1-\phi)^2}{\zeta(\phi)}\left[\frac{\nablu
    \cdot G(\phi)\tens{W}}{\phi}-\nablu\frac{\delta F}{\delta
    \phi}+\frac{2\,\nablu\cdot\,\phi\, \etam\,
    \Dm^{0}}{\phi}-\frac{2\nablu\cdot\,(1-\phi)\, \etas\,
    \Ds^{0}}{1-\phi}\right].
\label{eqn:relative}
\end{gather}
%
$G(\phi)\tens{W}$ is the viscoelastic  micellar
backbone stress, ${W}_{\alpha\beta}=\frac{\partial{R}'_{\alpha}}{\partial
  \vect{R}}\cdot\frac{\partial {R}'_{\beta}}{\partial
  \vect{R}}-\delta_{\alpha\beta}$ is the local strain that would have
to be reversed to relax this stress, and $G$ is the plateau modulus.
The free energy $F$ comprises osmotic and elastic parts:
  \be
  \label{eqn:free_energy}
  F =  \tfrac{1}{2} \int d^3q\, (1+\xi^2
  q^2)f''|\phi(q)|^2 + \tfrac{1}{2}\int d^3x G(\phi){\rm
  tr}[\tens{W}-\log(\tens{\delta}+\tens{W})], 
  \ee
  where $f''$ is the osmotic susceptibility and $\xi$ the equilibrium
  correlation length for concentration fluctuations. The Newtonian
  stress $2\phi\, \etam\, \Dm^{0}$ describes fast micellar
  processes (\eg\ Rouse modes) with $\Dm^{0}$ the traceless symmetric
  micellar strain rate tensor, and $2(1-\phi)\, \etas\,
  \Ds^{0}$ is the solvent stress.  Incompressibility determines the
  pressure $p$.  The drag coefficient $\zeta$ (Eq.~\ref{eqn:relative})
  ensures that the force $\zeta\vrel$ impedes relative motion.
  The micellar diffusion coefficient $D\propto f''/\zeta$. We have
  omitted negligible inertial corrections to Eqs.~(\ref{eqn:navier})
  and (\ref{eqn:relative}) \cite{long}.
  
  Eq.~(\ref{eqn:navier}) is the Navier Stokes equation generalised to
  include osmotic stresses.  Eq.~(\ref{eqn:relative}) is a generalised
  CH equation in which micelles diffuse in response to gradients in
  the osmotic force $\nablu [\delta F/\delta \phi]$ {\em and} in the
  viscoelastic stress; for $dG/d\phi >0$ (assumed here), HF feedback
  occurs.  The micellar strain obeys d-JS dynamics~\cite{johnson77}:
%
\begin{equation}
\label{eqn:JSd}
(\partial_t+\vm\cdot\nablu)\tens{W}=a(\Dm\cdot\tens{W}+\tens{W}\cdot\Dm)+
(\tens{W}\cdot\Omm-\Omm\cdot\tens{W})+2\Dm-\frac{\tens{W}}{\tau(\phi)}+
\frac{l^2}{\tau(\phi)}\nablu^2 \tens{W} 
\end{equation}
\end{widetext}
where $ 2\Omm=\nablu \vm - (\nablu \vm)^T$ with $(\nablu
\vm)_{\alpha\beta}\equiv \partial_{\alpha}(v_{\rm m})_\beta$.
$\tau(\phi)$ is the Maxwell time; $l$ is a length that could, for
example, be set by the mesh-size.  The slip parameter $a$ measures the
fractional stretch of the micelles compared to the flow.  For $|a|<1$
(slip) the intrinsic flow curve $\Sigma_{xy}(\gdot)$ is capable of the
non-monotonicity of Fig.~\ref{fig:schem}.

\begin{figure}[h]
\centering
\includegraphics[scale=0.25]{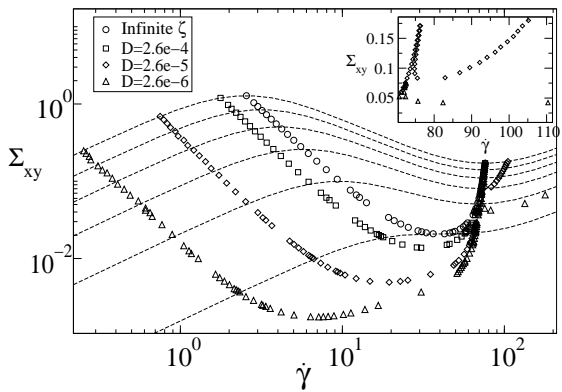}
\vskip-0.2truecm
\caption{Dashed lines: intrinsic flow curves for
  $\phi=0.11$, $0.091$, $0.072$,$0.053,0.034,0.015$ (downwards). Symbols:
  spinodals for the uncoupled limit $\zeta\to\infty$ ($\circ$);
  coupled model with $D(\phi=0.11)$ taken from DLS ($\square$);
  coupled model with artificially reduced $D$ 
  ($\lozenge$,$\vartriangle$).  Inset: zoom on large $\gdot$.
\label{fig:spinodals_with_phi}} 
\end{figure}

We study planar shear between infinite plates at $y=\{0,L\}$ with
$(\vect{v},\nablu v, \nablu \wedge \vect{v})$ in the
$(\hat{\vect{x}},\hat{\vect{y}},\hat{\vect{z}})$ directions. At the
plates we assume $\partial_y \phi=\partial^3_y \phi=0$, $\partial_y
\tens{W}=0$, and no slip.  For controlled strain rate (assumed
throughout) $\int_0^L dy \gdot(y)=\rm{const.}$ Unless stated, we
use model parameter values at $\phi=0.11$ from rheological data for
CTAB(0.3M)/${\rm NaNO}_3/{\rm H}_2{\rm O}$~\cite{lerouge_note}, and
light scattering (DLS) data for CTAB/KBr/H$_2$O~\cite{CanHirZan85}; we
calculate the drag as $\zeta=6\pi \bar{\eta}
\xi^{-2}$~\cite{pgdgpolymer} where
$\bar{\eta}=\phi\etam+(1-\phi)\etas$.  We extrapolate $G(\phi),\tau(\phi),
D(\phi),\zeta(\phi),\xi(\phi)$ to $\phi<0.11$ using scaling laws for
semidilute wormlike micelles.
We fix $a$ (assumed independent of $\phi$) by comparing with Cates'
model~\cite{cates90}; and use units in which $G(\phi=0.11)=1$,
$\tau(\phi=0.11)=1$ and $L=1$.  The homogeneous intrinsic flow curves
$\Sigma_{xy}(\gdot,\phi)$ that satisfy
$\partial_t\vv=\partial_t\phi=\partial_t\tens{W}=0$
are shown as dashed lines in Fig.~\ref{fig:spinodals_with_phi}.  The
region of negative slope ends at a ``critical'' point
$\phi_{\rm c}\approx 0.015$: CPCl/NaSal in brine~\cite{berret94b} shows the
same trend.

We linearise in fluctuations $\sum_{\vect{k}}[\delta \gdot,\delta
\tens{W},\delta\phi]_{\raisebox{-1mm}{\vect{k}}}\,e^{i\vect{k}\cdot\vect{r}
  + \omega t}$ about these homogeneous states. The stability analysis
$\omega_{\vect{k},\alpha}{\eigenvec} = \tens{M}_{\vect{k}}{\eigenvec}$
gives the normal modes ${\eigenvec}$ (where $\alpha$ is the mode
index), each encoding a set of relative amplitudes of
$\delta\gdot,\delta W_{ij},\delta \phi$. The lower spinodal lies where
the largest branch $\omega_{\vect{k}}$ of the dispersion relation
first goes positive (unstable) as the background homogeneous state is
swept up the intrinsic flow curve from $\gdot=0$; the upper spinodal
is defined likewise, for sweeps down from large $\gdot$.  We consider
only $\vect{k}=k\vect{\hat{y}}$; the stability of
$\vect{k}=k\vect{\hat{z}}$ is unaffected by shear in our model.

In the limit of infinite drag $\zeta\to\infty$ at fixed
$\nablu\cdot\frac{1}{\zeta}\nablu\frac{\delta F}{\delta \phi}$, the
coupling of $\delta\gdot$ and $\delta\tens{W}$ to $\delta\phi$ is
disabled.  Fluctuations in the mechanical subspace $[\delta
\gdot,\delta W_{xy},\delta W_{xx},\delta W_{yy}]$ then obey
conventional (uniform-$\phi$) d-JS dynamics, in which any homogeneous
shear state in the regime of negative constitutive slope
$d\Sigma_{xy}/d\gdot<0$ is unstable: the spinodal is given by circles
in Fig.~\ref{fig:spinodals_with_phi}.  The concentration fluctuations
independently obey conventional (zero-shear) CH dynamics, with a
demixing instability when $D<0$. We consider only {\em flow-induced}
instabilities here, $D>0$.
\vskip-0.2truecm

For finite drag, HF feedback couples these subspaces and can induce
instability even if $d\Sigma_{xy}/d\gdot>0$ and $D>0$. For model
parameters from the data of Refs.~\cite{lerouge_note,CanHirZan85}, the
mechanical instability is enhanced slightly by concentration coupling
(squares in Fig.~\ref{fig:spinodals_with_phi}). This effect
increases near a zero-shear demixing instability: see the diamonds and
triangles in Fig.~\ref{fig:spinodals_with_phi}, for smaller
$D(\phi=0.11)$.  (The lobe of instability at high shear rate is
discussed in Ref.~\cite{long}.)  Within some simplifying
assumptions~\cite{long}, a qualitative expression for the lower
spinodal is
%
$\tilde{D}\frac{d{\Sigma}_{xy}}{d{\gdot}}+\frac{dG}
{d\phi}\frac{{W}_{xy}}{\tilde{\zeta}}\frac{d{Z}}{d{\gdot}}=0$ 
%
where $2Z=(a-1)W_{xx}+(a+1)W_{yy}<0$,
$\tilde{\zeta}\propto\zeta$ and $\tilde{D}=D-\frac{dG}{d\phi}\frac{Z}{\zeta}$.

Enhancement of flow instabilities by positive feedback with
concentration was first predicted by the remarkable insight of Schmitt
{\it et al.}\cite{schmitt95}. However they directly assumed a chemical
potential $\mu=\mu(\gdot)$. Although this is equivalent to our
approach in the limit of adiabatic stress response (assumed in
\cite{schmitt95}), below we find that the dynamics inside the spinodal
{\em are} dictated by the micellar stress response.  The spinodal is
unaffected since the dynamics are adiabatic here by definition: the
above condition corresponds to Schmitt's Eq.~(24).

Experimentally the spinodal is found via sweeps along the intrinsic
flow curve, though in practice banding can occur prematurely via
metastable kinetics~\cite{Berr97}.  The same ambiguity arises in the
spinodal of conventional fluid-vapour demixing, defined via
quasistatic compression.

\begin{figure*}[htbp]
  \centering \subfigure[$\zeta\to\infty,\,\gdot=7.0$]{
    \includegraphics[scale=0.32]{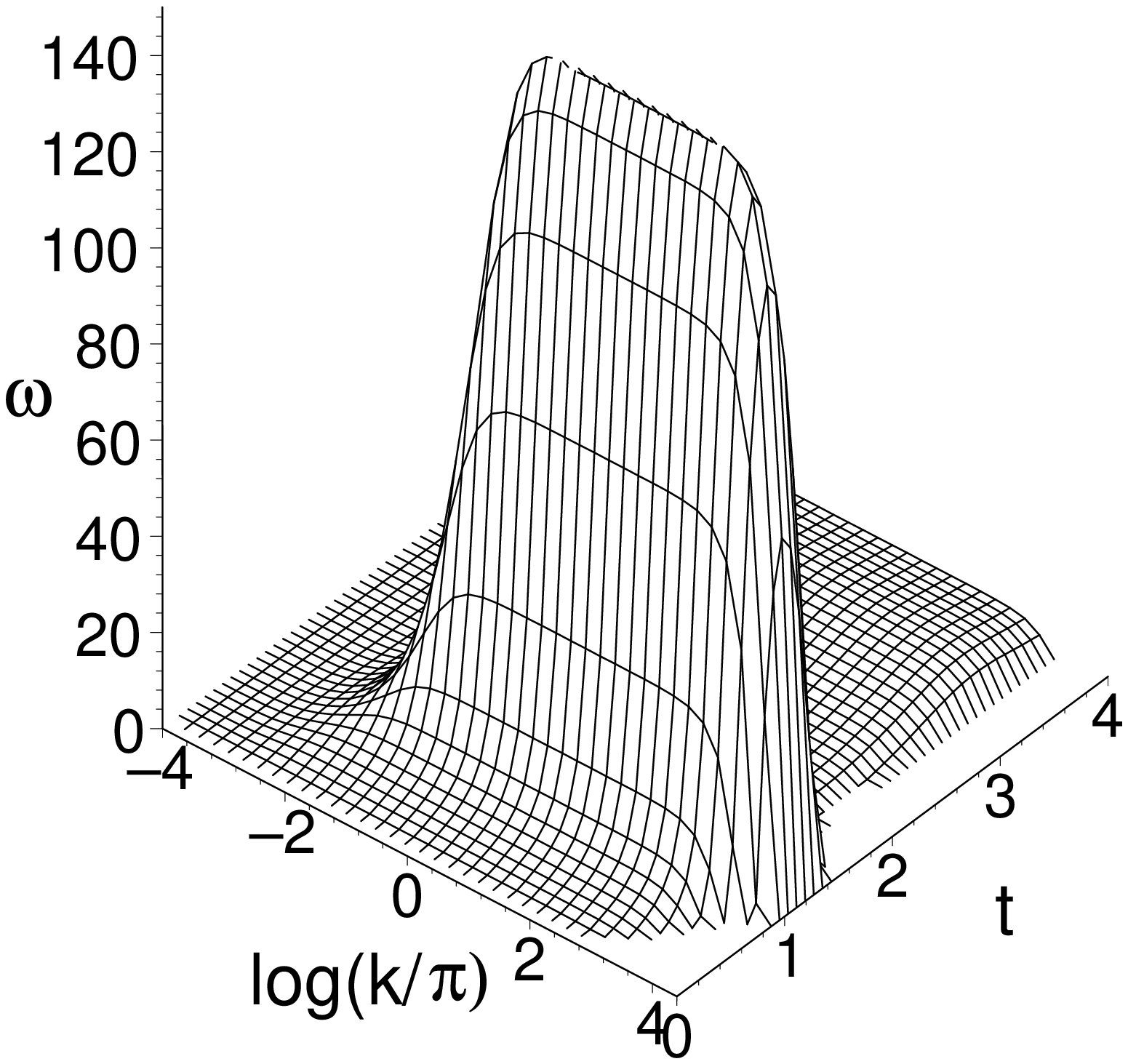}
\label{fig:dispersion:a}
}\hspace{0.2cm}
\subfigure[\scriptsize $D_{\phi=0.11}=2.6\times 10^{-4},\,\gdot=4.0$]{
\includegraphics[scale=0.32]{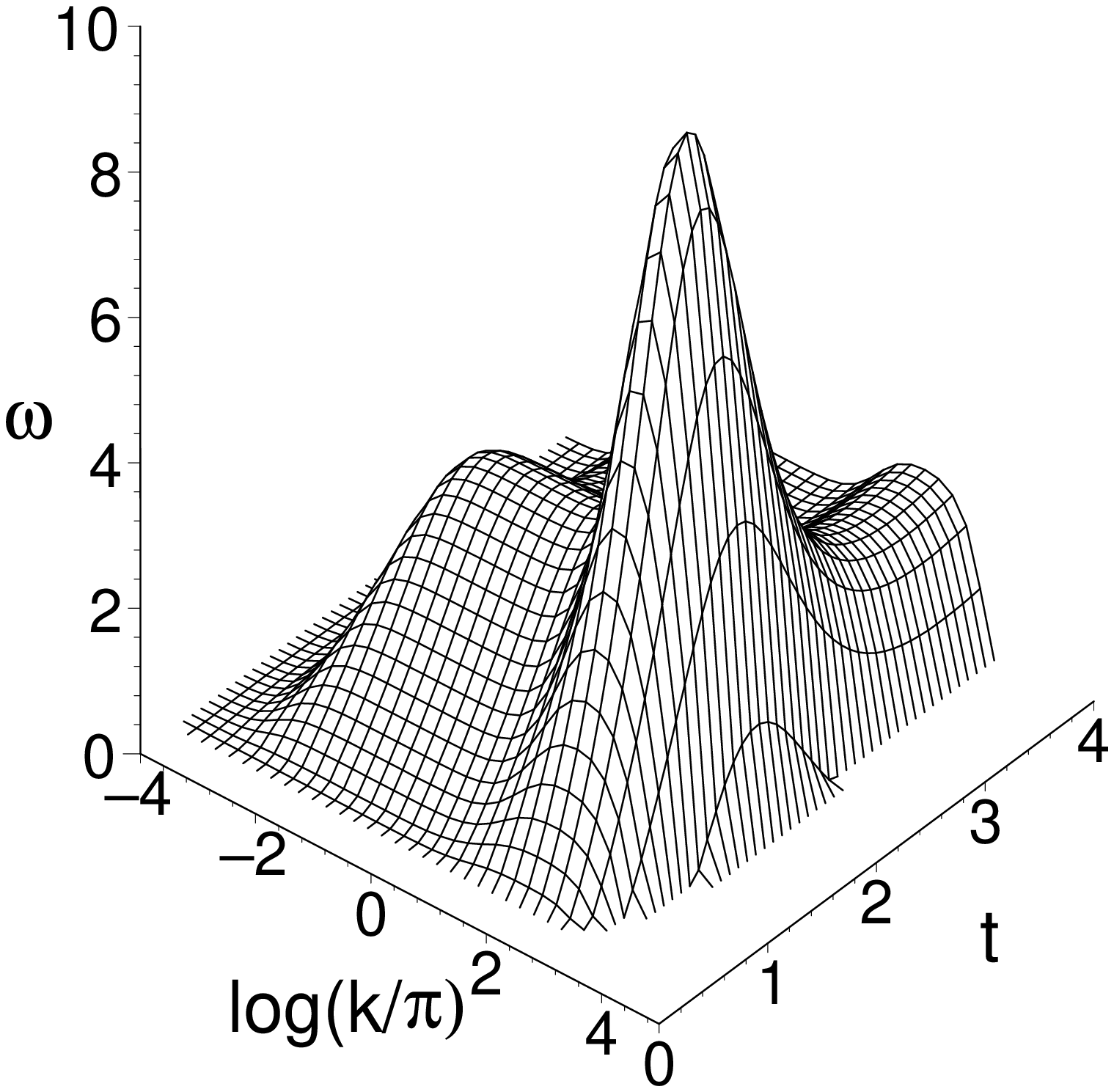}
\label{fig:dispersion:b}
}\hspace{0.2cm}
\subfigure[\scriptsize $D_{\phi=0.11}=2.6\times 10^{-9},\,\gdot=0.01$]{
\includegraphics[scale=0.32]{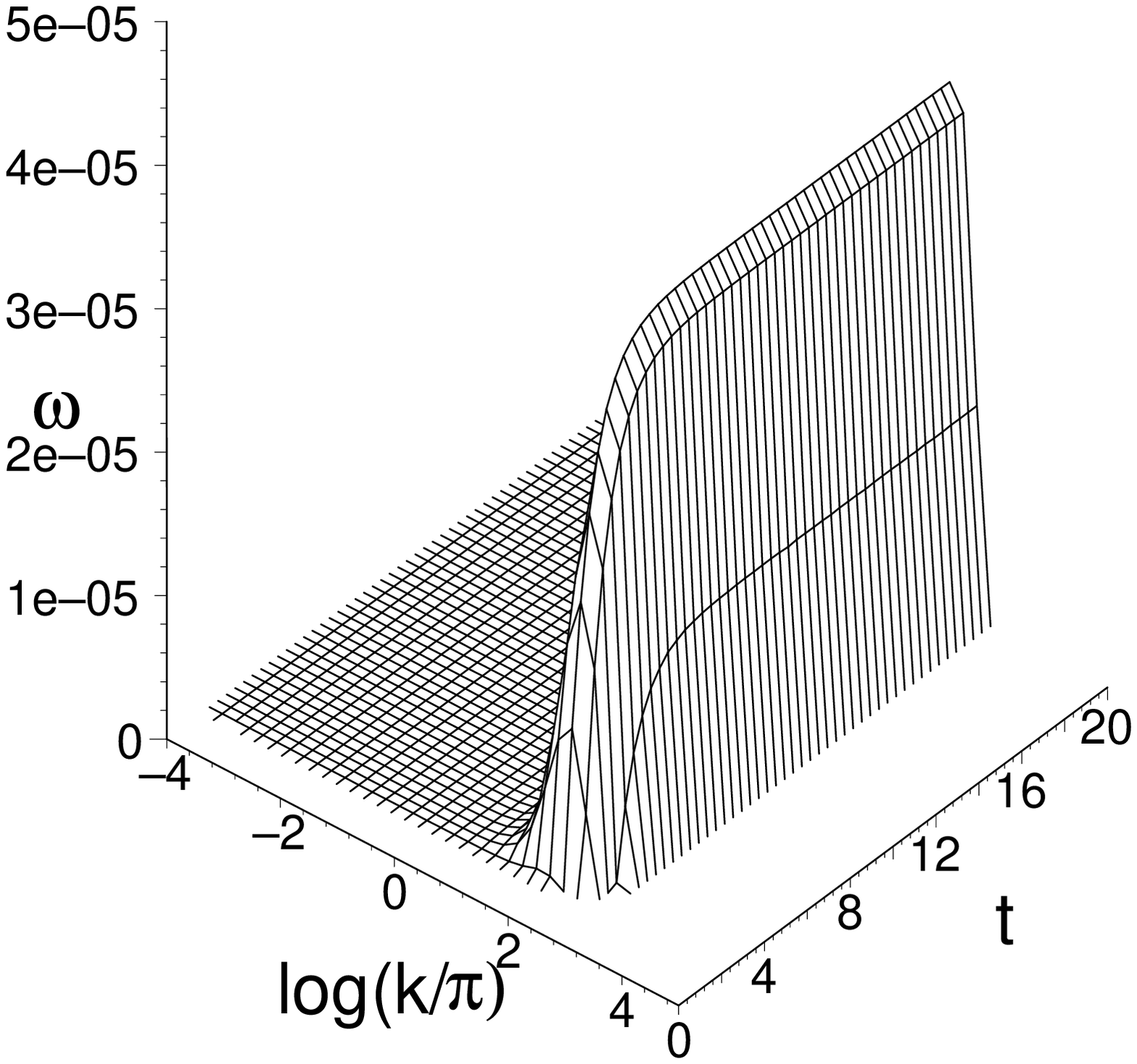}
\label{fig:dispersion:c}
}
\\\vspace{-0.5cm}
\subfigure{ 
  \includegraphics[scale=0.18]{vecmax_t_raw_phi0.11_gdot7.0.eps}
}
\subfigure{ 
  \includegraphics[scale=0.18]{vecmax_t_D2.6e-4_phi0.11_gdot4.0.eps}
}
\subfigure{ 
  \includegraphics[scale=0.18]{vecmax_t_D2.6e-9_phi0.11_gdot0.01.eps}
}
\vspace{-0.5cm}
\caption{Startup dispersion relation $\omega_k$ (top) and eigenvector at max. in $\omega_k$ (bottom). $Z=\tfrac{1}{2}(a-1)W_{xx}+\tfrac{1}{2}(1+a)W_{yy}$; $Y=\tfrac{1}{2}(1-a)W_{xx}-\tfrac{1}{2}(1+a)W_{yy}$.}
\label{fig:dispersion}
\vspace{-0.5cm} 
\end{figure*}

We now study the early-time kinetics in the unstable regime.  We
cannot assume that the system starts in a homogeneous state on the
unstable intrinsic flow curve, since the Maxwell time $\tau$ needed to
prepare such a state is longer than the typical time scales
$1/\omega_k$ of the instability itself (except very near the
spinodal)~\footnote{For temperature quenches in CH demixing, the system
essentially starts on the unstable branch of the chemical potential
because heat-conduction is fast enough to effect the quench before
demixing starts.}, and so
\begin{figure}[h]
\centerline{\psfig{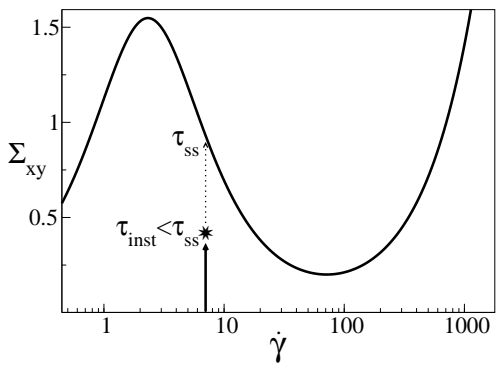}}
\vspace{-0.3cm}
\caption{Instability occurring as the homogeneous startup flow evolves
  towards the intrinsic unstable constitutive curve. 
\label{fig:schematic_stress_evolution} } 
\vspace{-0.2cm}
\end{figure}
most startup flows go unstable before the intrinsic flow curve can be
attained.  When the rheometer plate is first set moving the shear rate
rapidly homogenises, on the Reynolds timescale $\rho L^2/\eta\ll
\tau$.  The shear stress $\Sigma(t)=G\tens{W}(t)+\eta\gdot$ then {\em
  starts} to evolve towards the intrinsic flow curve, with
$\tens{W}(t)$ {\em initially} given by the homogeneous startup
solution of Eq.~(\ref{eqn:JSd}).  Without noise, it would attain
this flow curve at some time $\tau_{\rm ss}=O(\tau)$.  But of course
the system
has noise. The stability problem is now time-dependent:
$\omega_{k,\alpha}(t){\eigenvec}(t)=\tens{M}_{k}(t){\eigenvec}(t)$
since $\tens{M}_{k}\hspace{-0.2cm} =
\hspace{-0.15cm}\tens{M}_{k}({\gdot}, {\tens{W}}(t),{\phi})$.  At
first all dispersion branches are stable. Then at some time
$t_0\le\tau_{\rm ss}$ a branch goes unstable. The size of the growing
fluctuations at $t>t_0$ is $A_{k}(t)\sim \exp\left[\int_{t_0}^{t} dt'
  \omega_{k}(t')\right]$.  A rough criterion for detectability is
$\log A=O(10)$, which defines a $k-$dependent time scale $\tau_{\rm
  inst}(k)$ via $\int_{t_0}^{\tau_{\rm inst}(k)} dt'
\omega_{k}(t')=O(10)$. In most regimes, fluctuations emerge fastest at
a selected wavevector $k^*$, because of a peak in the dispersion
relation $\omega_{k}(t)$ (Fig.~\ref{fig:dispersion}).  We thus define
the overall time scale of the instability to be $\tau_{\rm
  inst}=\tau_{\rm inst}(k^*)$
(Fig.~\ref{fig:schematic_stress_evolution}).

Fig.~\ref{fig:dispersion} gives the unstable dispersion branch
$\omega_k(t)$ and the eigenvector at the maximum in $\omega_k$ for
three startup flows. $\tau_{\rm inst}$ is marked in each case. (We
used a continuous $k$ axis though in practice only harmonics $k=n\pi$
are allowed.)  For the pure mechanical instability
($\zeta\rightarrow\infty$, Fig.~\ref{fig:dispersion:a}), $\omega_{k}$
is cut off at large $k$ by interfaces and at low $k$ (beyond the gap
size) by Reynolds effects.  Between there is a plateau where
$\omega_{k}$ is set by the viscoelastic stress response, with no
clearly selected lengthscale.  Fig.~\ref{fig:dispersion:b} is for a
coupled model far from a zero-shear CH demixing instability (with the
spinodal given by circles in Fig.~\ref{fig:spinodals_with_phi}).
Concentration coupling has enhanced the mechanical instability at
short length scales; in competition with the interfacial terms this
selects a length scale $k^{*-1}$, as seen
experimentally~\cite{DecLerBer01}. The eigenvector at $k^*$
(Fig.~\ref{fig:dispersion:b}, bottom) is mainly still in the $\delta\gdot$
and $\delta\tens{W}$ directions, as expected. The plateau of the mechanical
instability is still apparent at smaller $k$, since this shear rate
would have been unstable even for $\zeta\to\infty$.  Concentration
diffusion cannot keep pace at these length scales, and is absent from
the eigenvector.  Fig.~\ref{fig:dispersion:c} is for a system close to
zero-shear demixing, for which the lower spinodal is at a shear rate
far below the spinodal of the pure mechanical instability (triangles
in Fig.~\ref{fig:spinodals_with_phi}). Our imposed shear rate is just
inside this spinodal, so the mechanical plateau is absent from the
dispersion relation, leaving only the diffusive concentration-coupled
branch.  The eigenvector is dominated by $\delta\phi$, so this
instability is essentially demixing, triggered by flow. It occurs
slowly enough that the intrinsic flow curve is reached before
fluctuations grow appreciably: $\tau_{\rm inst}$ is off the scale of
Fig.~\ref{fig:dispersion:c}.

In summary, we have studied the early time kinetics of the shear
banding instability in the d-JS model with 2-fluid coupling to
concentration. The spinodal onset of pure mechanical
instability with negative constitutive slope $d\Sigma_{xy}/d\gdot<0$
(mathematically indistinguishable from a flow-induced transition in,
for example, nematic liquid crystals \cite{olmstedlu99}) is shifted by
coupling to concentration. For startup quenches deep in the unstable
region the instability in general occurs before the homogeneous
startup flow can attain the intrinsic flow curve. An initial length
scale is selected only if the instability is coupled to concentration.
These results are qualitatively consistent with recent experiments on
wormlike micellar solutions~\cite{lerouge_note}.

{\bf Acknowledgments} We thank Paul Callaghan, Ron Larson, Sandra
Lerouge, and Tom McLeish for interesting discussions and EPSRC GR/N
11735 for financial support; this work was supported in part by the
National Science Foundation under Grant No.PHY99-07949.

\begin{thebibliography}{50}
\expandafter\ifx\csname natexlab\endcsname\relax\def\natexlab#1{#1}\fi
\expandafter\ifx\csname bibnamefont\endcsname\relax
  \def\bibnamefont#1{#1}\fi
\expandafter\ifx\csname bibfnamefont\endcsname\relax
  \def\bibfnamefont#1{#1}\fi
\expandafter\ifx\csname citenamefont\endcsname\relax
  \def\citenamefont#1{#1}\fi
\expandafter\ifx\csname url\endcsname\relax
  \def\url#1{\texttt{#1}}\fi
\expandafter\ifx\csname urlprefix\endcsname\relax\def\urlprefix{URL }\fi
\providecommand{\bibinfo}[2]{#2}
\providecommand{\eprint}[2][]{\url{#2}}

\bibitem[{\citenamefont{Cates}(1990)}]{cates90}
\bibinfo{author}{\bibfnamefont{M.~E.} \bibnamefont{Cates}},
  \bibinfo{journal}{J.~Phys. Chem.} \textbf{\bibinfo{volume}{94}},
  \bibinfo{pages}{371} (\bibinfo{year}{1990}).

\bibitem[{\citenamefont{Spenley and Cates}(1994)}]{SpenCate94}
\bibinfo{author}{\bibfnamefont{N.~A.} \bibnamefont{Spenley}} \bibnamefont{and}
  \bibinfo{author}{\bibfnamefont{M.~E.} \bibnamefont{Cates}},
  \bibinfo{journal}{Macromolecules} \textbf{\bibinfo{volume}{27}},
  \bibinfo{pages}{3850} (\bibinfo{year}{1994}). 
\bibinfo{author}{\bibfnamefont{N.~A.} \bibnamefont{Spenley}},
\bibinfo{author}{\bibfnamefont{X.-F.} \bibnamefont{Yuan}},
 \bibnamefont{and}
  \bibinfo{author}{\bibfnamefont{M.~E.} \bibnamefont{Cates}},
  \bibinfo{journal}{J. Phys. II (Fr)} \textbf{\bibinfo{volume}{6}},
  \bibinfo{pages}{551} (\bibinfo{year}{1996}). 

\bibitem[{\citenamefont{Yerushalmi et~al.}(1970)\citenamefont{Yerushalmi, Katz,
  and Shinnar}}]{Yerushalmi70}
\bibinfo{author}{\bibfnamefont{J.}~\bibnamefont{Yerushalmi}},
  \bibinfo{author}{\bibfnamefont{S.}~\bibnamefont{Katz}}, \bibnamefont{and}
  \bibinfo{author}{\bibfnamefont{R.}~\bibnamefont{Shinnar}},
  \bibinfo{journal}{Chemical Engineering Science}
  \textbf{\bibinfo{volume}{25}}, \bibinfo{pages}{1891} (\bibinfo{year}{1970}).

\bibitem[{\citenamefont{Spenley et~al.}(1993)\citenamefont{Spenley, Cates, and
  McLeish}}]{SCM93}
\bibinfo{author}{\bibfnamefont{N.~A.} \bibnamefont{Spenley}},
  \bibinfo{author}{\bibfnamefont{M.~E.} \bibnamefont{Cates}}, \bibnamefont{and}
  \bibinfo{author}{\bibfnamefont{T.~C.~B.} \bibnamefont{McLeish}},
  \bibinfo{journal}{Phys. Rev. Lett.} \textbf{\bibinfo{volume}{71}},
  \bibinfo{pages}{939} (\bibinfo{year}{1993}).

\bibitem[{\citenamefont{Dhont}(1999)}]{Dhon99}
\bibinfo{author}{\bibfnamefont{J.~K.~G.} \bibnamefont{Dhont}},
  \bibinfo{journal}{Phys. Rev. {\bf E}} \textbf{\bibinfo{volume}{60}},
  \bibinfo{pages}{4534} (\bibinfo{year}{1999}).
  \bibinfo{author}{\bibfnamefont{X.-F.}~\bibnamefont{Yuan}},
  \bibinfo{journal}{Europhys. Lett.} \textbf{\bibinfo{volume}{46}},
  \bibinfo{pages}{542} (\bibinfo{year}{1999}{\natexlab{a}}).

\bibitem[{\citenamefont{Rehage and Hoffmann}(1991)}]{rehage}
\bibinfo{author}{\bibfnamefont{H.}~\bibnamefont{Rehage}} \bibnamefont{and}
  \bibinfo{author}{\bibfnamefont{H.}~\bibnamefont{Hoffmann}},
  \bibinfo{journal}{Mol. Phys.} \textbf{\bibinfo{volume}{74}},
  \bibinfo{pages}{933} (\bibinfo{year}{1991}).

\bibitem[{\citenamefont{Berret et~al.}(1994{\natexlab{a}})\citenamefont{Berret,
  Roux, Porte, and Lindner}}]{SANS}
\bibinfo{author}{\bibfnamefont{J.~F.} \bibnamefont{Berret}},
  \bibinfo{author}{\bibfnamefont{D.~C.} \bibnamefont{Roux}},
  \bibinfo{author}{\bibfnamefont{G.}~\bibnamefont{Porte}}, \bibnamefont{and}
  \bibinfo{author}{\bibfnamefont{P.}~\bibnamefont{Lindner}},
  \bibinfo{journal}{Europhys. Lett.} \textbf{\bibinfo{volume}{25}},
  \bibinfo{pages}{521} (\bibinfo{year}{1994}{\natexlab{a}}).
\bibinfo{author}{\bibfnamefont{V.}~\bibnamefont{Schmitt}},
  \bibinfo{author}{\bibfnamefont{F.}~\bibnamefont{Lequeux}},
  \bibinfo{author}{\bibfnamefont{A.}~\bibnamefont{Pousse}}, \bibnamefont{and}
  \bibinfo{author}{\bibfnamefont{D.}~\bibnamefont{Roux}},
  \bibinfo{journal}{Langmuir} \textbf{\bibinfo{volume}{10}},
  \bibinfo{pages}{955} (\bibinfo{year}{1994}).

\bibitem[{\citenamefont{Berret et~al.}(1994{\natexlab{b}})\citenamefont{Berret,
  Roux, and Porte}}]{berret94b}
\bibinfo{author}{\bibfnamefont{J.~F.} \bibnamefont{Berret}},
  \bibinfo{author}{\bibfnamefont{D.~C.} \bibnamefont{Roux}}, \bibnamefont{and}
  \bibinfo{author}{\bibfnamefont{G.}~\bibnamefont{Porte}},
  \bibinfo{journal}{J.~Phys.~II (France)} \textbf{\bibinfo{volume}{4}},
  \bibinfo{pages}{1261} (\bibinfo{year}{1994}{\natexlab{b}}).

\bibitem[{\citenamefont{Decruppe et~al.}(1995)\citenamefont{Decruppe, Cressely,
  Makhloufi, and Cappelaere}}]{FB}
\bibinfo{author}{\bibfnamefont{R.}~\bibnamefont{Makhloufi}},
  \bibinfo{author}{\bibfnamefont{J.~P.} \bibnamefont{Decruppe}},
  \bibinfo{author}{\bibfnamefont{A.}~\bibnamefont{Aitali}}, \bibnamefont{and}
  \bibinfo{author}{\bibfnamefont{R.}~\bibnamefont{Cressely}},
  \bibinfo{journal}{Europhys. Lett.} \textbf{\bibinfo{volume}{32}},
  \bibinfo{pages}{253} (\bibinfo{year}{1995}).
\bibinfo{author}{\bibfnamefont{J.~P.} \bibnamefont{Decruppe}},
  \bibinfo{author}{\bibfnamefont{E.}~\bibnamefont{Cappelaere}},
  \bibnamefont{and} \bibinfo{author}{\bibfnamefont{R.}~\bibnamefont{Cressely}},
  \bibinfo{journal}{J. Phys. II (France)} \textbf{\bibinfo{volume}{7}},
  \bibinfo{pages}{257} (\bibinfo{year}{1997}).
\bibinfo{author}{\bibfnamefont{J.~F.} \bibnamefont{Berret}},
  \bibinfo{author}{\bibfnamefont{G.}~\bibnamefont{Porte}}, \bibnamefont{and}
  \bibinfo{author}{\bibfnamefont{J.~P.} \bibnamefont{Decruppe}},
  \bibinfo{journal}{Phys. Rev. {\bf E}} \textbf{\bibinfo{volume}{55}},
  \bibinfo{pages}{1668} (\bibinfo{year}{1997}).

\bibitem[{\citenamefont{Callaghan et~al.}(1996)\citenamefont{Callaghan, Cates,
  Rofe, and Smeulders}}]{Call+96}
\bibinfo{author}{\bibfnamefont{R.~W.} \bibnamefont{Mair}} \bibnamefont{and}
  \bibinfo{author}{\bibfnamefont{P.~T.} \bibnamefont{Callaghan}},
  \bibinfo{journal}{Europhys. Lett.} \textbf{\bibinfo{volume}{36}},
  \bibinfo{pages}{719} (\bibinfo{year}{1996}{\natexlab{a}});
  \bibinfo{journal}{J. Rheol} \textbf{\bibinfo{volume}{41}},
  \bibinfo{pages}{901} (\bibinfo{year}{1997}{\natexlab{b}}).
\bibinfo{author}{\bibfnamefont{M.~M.} \bibnamefont{Britton}} \bibnamefont{and}
  \bibinfo{author}{\bibfnamefont{P.~T.} \bibnamefont{Callaghan}},
  \bibinfo{journal}{Phys. Rev. Lett.} \textbf{\bibinfo{volume}{78}},
  \bibinfo{pages}{4930} (\bibinfo{year}{1997}).
\bibitem[{\citenamefont{Berret et~al.}(1998)\citenamefont{Berret, Roux, and
  Lindner}}]{BRL98}
\bibinfo{author}{\bibfnamefont{J.~F.} \bibnamefont{Berret}},
  \bibinfo{author}{\bibfnamefont{D.~C.} \bibnamefont{Roux}}, \bibnamefont{and}
  \bibinfo{author}{\bibfnamefont{P.}~\bibnamefont{Lindner}},
  \bibinfo{journal}{European Physical Journal B} \textbf{\bibinfo{volume}{5}},
  \bibinfo{pages}{67} (\bibinfo{year}{1998}).
\bibitem[{\citenamefont{Olmsted et~al.}(2000)\citenamefont{Olmsted, Radulescu,
  and Lu}}]{olmsted99a}
\bibinfo{author}{\bibfnamefont{C.-Y.~D.} \bibnamefont{Lu}},
  \bibinfo{author}{\bibfnamefont{P.~D.} \bibnamefont{Olmsted}},
  \bibnamefont{and} \bibinfo{author}{\bibfnamefont{R.~C.} \bibnamefont{Ball}},
  \bibinfo{journal}{Phys. Rev. Lett.} \textbf{\bibinfo{volume}{84}},
  \bibinfo{pages}{642} (\bibinfo{year}{2000}).
\bibinfo{author}{\bibfnamefont{P.~D.} \bibnamefont{Olmsted}} \bibnamefont{and}
  \bibinfo{author}{\bibfnamefont{P.~M.} \bibnamefont{Goldbart}},
  \bibinfo{journal}{Phys.~Rev.} \textbf{\bibinfo{volume}{A46}},
  \bibinfo{pages}{4966} (\bibinfo{year}{1992}).
\bibinfo{author}{\bibfnamefont{J.~R.~A.} \bibnamefont{Pearson}},
  \bibinfo{journal}{J. Rheol.} \textbf{\bibinfo{volume}{38}},
  \bibinfo{pages}{309} (\bibinfo{year}{1994}).

\bibitem[{\citenamefont{Berret}(1997)}]{Berr97}
\bibinfo{author}{\bibfnamefont{J.~F.} \bibnamefont{Berret}},
  \bibinfo{journal}{Langmuir} \textbf{\bibinfo{volume}{13}},
  \bibinfo{pages}{2227} (\bibinfo{year}{1997}).
\bibinfo{author}{\bibfnamefont{C.}~\bibnamefont{Grand}},
  \bibinfo{author}{\bibfnamefont{J.}~\bibnamefont{Arrault}}, \bibnamefont{and}
  \bibinfo{author}{\bibfnamefont{M.~E.} \bibnamefont{Cates}},
  \bibinfo{journal}{J.~Phys.~II (France)} \textbf{\bibinfo{volume}{7}},
  \bibinfo{pages}{1071} (\bibinfo{year}{1997}).
\bibinfo{author}{\bibfnamefont{S.}~\bibnamefont{Lerouge}},
  \bibinfo{author}{\bibfnamefont{J.~P.} \bibnamefont{Decruppe}},
  \bibnamefont{and} \bibinfo{author}{\bibfnamefont{C.}~\bibnamefont{Humbert}},
  \bibinfo{journal}{Phys. Rev. Lett.} \textbf{\bibinfo{volume}{81}},
  \bibinfo{pages}{5457} (\bibinfo{year}{1998}).
\bibinfo{author}{\bibfnamefont{S.}~\bibnamefont{Lerouge}},
  \bibinfo{author}{\bibfnamefont{J.~P.} \bibnamefont{Decruppe}},
  \bibnamefont{and} \bibinfo{author}{\bibfnamefont{J.~F.}
  \bibnamefont{Berret}}, \bibinfo{journal}{Langmuir}
  \textbf{\bibinfo{volume}{16}}, \bibinfo{pages}{6464} (\bibinfo{year}{2000}).
\bibinfo{author}{\bibfnamefont{J.~F.} \bibnamefont{Berret}} \bibnamefont{and}
  \bibinfo{author}{\bibfnamefont{G.}~\bibnamefont{Porte}},
  \bibinfo{journal}{Phys. Rev. {\bf E}} \textbf{\bibinfo{volume}{60}},
  \bibinfo{pages}{4268} (\bibinfo{year}{1999}).

\bibitem[{\citenamefont{Decruppe et~al.}(2001)\citenamefont{Decruppe, Lerouge,
  and Berret}}]{DecLerBer01}
\bibinfo{author}{\bibfnamefont{J.~P.} \bibnamefont{Decruppe}},
  \bibinfo{author}{\bibfnamefont{S.}~\bibnamefont{Lerouge}}, \bibnamefont{and}
  \bibinfo{author}{\bibfnamefont{J.~F.} \bibnamefont{Berret}},
  \bibinfo{journal}{Phys.\ Rev.\ E} \textbf{\bibinfo{volume}{6302}},
  \bibinfo{pages}{2501} (\bibinfo{year}{2001}).

\bibitem[{\citenamefont{Helfand and Fredrickson}(1989)}]{HelfFred89}
\bibinfo{author}{\bibfnamefont{E.}~\bibnamefont{Helfand}} \bibnamefont{and}
  \bibinfo{author}{\bibfnamefont{G.~H.} \bibnamefont{Fredrickson}},
  \bibinfo{journal}{Phys. Rev. Lett.} \textbf{\bibinfo{volume}{62}},
  \bibinfo{pages}{2468} (\bibinfo{year}{1989}).

\bibitem[{\citenamefont{Brochard and de~Gennes}(1977)}]{brochdgen77}
\bibinfo{author}{\bibfnamefont{P.-G.} \bibnamefont{de~Gennes}},
  \bibinfo{journal}{Macromolecules} \textbf{\bibinfo{volume}{9}},
  \bibinfo{pages}{587} (\bibinfo{year}{1976}).
\bibinfo{author}{\bibfnamefont{F.}~\bibnamefont{Brochard}},
  \bibinfo{journal}{J.Phys. (Paris)} \textbf{\bibinfo{volume}{44}},
  \bibinfo{pages}{39} (\bibinfo{year}{1983}).

\bibitem[{\citenamefont{Doi and Onuki}(1992)}]{DoiOnuk92}
\bibinfo{author}{\bibfnamefont{M.}~\bibnamefont{Doi}} \bibnamefont{and}
  \bibinfo{author}{\bibfnamefont{A.}~\bibnamefont{Onuki}}, \bibinfo{journal}{J.
  Phys. II (France)} \textbf{\bibinfo{volume}{2}}, \bibinfo{pages}{1631}
  (\bibinfo{year}{1992}).

\bibitem[{\citenamefont{Milner}(1993)}]{milner93}
\bibinfo{author}{\bibfnamefont{S.~T.} \bibnamefont{Milner}},
  \bibinfo{journal}{Phys.~Rev.} \textbf{\bibinfo{volume}{E48}},
  \bibinfo{pages}{3674} (\bibinfo{year}{1993});
\bibinfo{journal}{Phys. Rev. Lett.} \textbf{\bibinfo{volume}{66}},
\bibinfo{pages}{1477} (\bibinfo{year}{1991}).

\bibitem[{\citenamefont{Wu et~al.}(1991)\citenamefont{Wu, Pine, and
  Dixon}}]{WPD91}
\bibinfo{author}{\bibfnamefont{X.~L.} \bibnamefont{Wu}},
  \bibinfo{author}{\bibfnamefont{D.~J.} \bibnamefont{Pine}}, \bibnamefont{and}
  \bibinfo{author}{\bibfnamefont{P.~K.} \bibnamefont{Dixon}},
  \bibinfo{journal}{Phys. Rev. Lett.} \textbf{\bibinfo{volume}{66}},
  \bibinfo{pages}{2408} (\bibinfo{year}{1991}).

\bibitem[{\citenamefont{Clarke and McLeish}(1998)}]{ClarMcle98}
\bibinfo{author}{\bibfnamefont{N.}~\bibnamefont{Clarke}} \bibnamefont{and}
  \bibinfo{author}{\bibfnamefont{T.~C.~B.} \bibnamefont{McLeish}},
  \bibinfo{journal}{Phys. Rev. {\bf E}} \textbf{\bibinfo{volume}{57}},
  \bibinfo{pages}{R3731} (\bibinfo{year}{1998}).
\bibinfo{author}{\bibfnamefont{T.}~\bibnamefont{Sun}} \bibnamefont{and}
  \bibinfo{author}{\bibfnamefont{A.~C.} \bibnamefont{Balazs}} \bibnamefont{and} \bibinfo{author}{\bibfnamefont{D.} \bibnamefont{Jasnow}},
  \bibinfo{journal}{Phys. Rev. {\bf E}} \textbf{\bibinfo{volume}{55}},
  \bibinfo{pages}{R6344} (\bibinfo{year}{1997}).

\bibitem[{lon()}]{long}
\bibinfo{note}{S M Fielding and P D Olmsted. In preparation.}

\bibitem[{\citenamefont{Johnson and Segalman}(1977)}]{johnson77}
\bibinfo{author}{\bibfnamefont{M.}~\bibnamefont{Johnson}} \bibnamefont{and}
  \bibinfo{author}{\bibfnamefont{D.}~\bibnamefont{Segalman}},
  \bibinfo{journal}{J. Non-Newt. Fl. Mech} \textbf{\bibinfo{volume}{2}},
  \bibinfo{pages}{255} (\bibinfo{year}{1977}).
\bibinfo{author}{\bibfnamefont{P.~D.} \bibnamefont{Olmsted}},
  \bibinfo{author}{\bibfnamefont{O.}~\bibnamefont{Radulescu}},
  \bibnamefont{and} \bibinfo{author}{\bibfnamefont{C.-Y.~D.} \bibnamefont{Lu}},
  \bibinfo{journal}{J. Rheology} \textbf{\bibinfo{volume}{44}},
  \bibinfo{pages}{257} (\bibinfo{year}{2000}).

\bibitem[{\citenamefont{de~Gennes}(1975)}]{pgdgpolymer}
\bibinfo{author}{\bibfnamefont{P.~G.} \bibnamefont{de~Gennes}},
  \emph{\bibinfo{title}{Scaling Concepts in Polymer Physics}}
  (\bibinfo{publisher}{Cornell}, \bibinfo{address}{Ithaca},
  \bibinfo{year}{1975}).

\bibitem[{ler()}]{lerouge_note}
\bibinfo{note}{S Lerouge. PhD Thesis, University of Metz, 2000.}

\bibitem[{\citenamefont{Candau et~al.}(1985)\citenamefont{Candau, Hirsch, and
  Zana}}]{CanHirZan85}
\bibinfo{author}{\bibfnamefont{S.~J.} \bibnamefont{Candau}},
  \bibinfo{author}{\bibfnamefont{E.}~\bibnamefont{Hirsch}}, \bibnamefont{and}
  \bibinfo{author}{\bibfnamefont{R.}~\bibnamefont{Zana}}, \bibinfo{journal}{J.\
  Colloid Interface Sci.} \textbf{\bibinfo{volume}{105}}, \bibinfo{pages}{521}
  (\bibinfo{year}{1985}).




\bibitem[{\citenamefont{Schmitt et~al.}(1995)\citenamefont{Schmitt, Marques,
  and Lequeux}}]{schmitt95}
\bibinfo{author}{\bibfnamefont{V.}~\bibnamefont{Schmitt}},
  \bibinfo{author}{\bibfnamefont{C.~M.} \bibnamefont{Marques}},
  \bibnamefont{and} \bibinfo{author}{\bibfnamefont{F.}~\bibnamefont{Lequeux}},
  \bibinfo{journal}{Phys.~Rev.} \textbf{\bibinfo{volume}{E52}},
  \bibinfo{pages}{4009} (\bibinfo{year}{1995}).
\bibitem[{\citenamefont{Olmsted and Lu}(1999)}]{olmstedlu99}
\bibinfo{author}{\bibfnamefont{P.~D.} \bibnamefont{Olmsted}} \bibnamefont{and}
  \bibinfo{author}{\bibfnamefont{C.-Y.~D.} \bibnamefont{Lu}},
  \bibinfo{journal}{Phys.~Rev.} \textbf{\bibinfo{volume}{E60}},
  \bibinfo{pages}{4397} (\bibinfo{year}{1999}).


\end{thebibliography}

\end{document}